\documentclass[aps,prapplied,twocolumn,showpacs,superscriptaddress]{revtex4-2}
\usepackage{graphicx}
\usepackage{amsmath}
\usepackage{amssymb}
\usepackage[normalem]{ulem}
\usepackage{float}
\usepackage[table]{xcolor}
\usepackage{tabularx}
\usepackage{array}

\newcommand{\e}{\textup{e}}
\newcommand{\ii}{\textup{i}}

\graphicspath{{converted_graphics/}}

\begin{document}
\title{Experimental proposal of a mode sorter for vector vortex beams of arbitrary order}

\author{P. Schneider Lacerda}
\affiliation{Departamento de F\'isica, Universidade Federal de Santa Catarina, Florian\'opolis, SC, 88040-900, Brazil}

\author{L. Marques Fagundes}
\affiliation{Departamento de F\'isica, Universidade Federal de Santa Catarina, Florian\'opolis, SC, 88040-900, Brazil}

\author{R. Medeiros de Ara\'ujo}
\affiliation{Departamento de F\'isica, Universidade Federal de Santa Catarina, Florian\'opolis, SC, 88040-900, Brazil}
\email{renne.araujo@ufsc.br}

\date{\today}
\begin{abstract}
We propose an experimental scheme for sorting vector vortex beams using a system of three optical cavities. The method targets the separation of the four vector vortex modes of a given order $m$ by exploiting their symmetry properties with respect to radial axes. The first cavity separates the modes into two pairs -- one symmetric and one antisymmetric under horizontal flip -- by transmitting one pair and reflecting the other. Each output is then directed into a second cavity, which further resolves the pairs into individual modes based on their parity. A key element between the cavities is a novel system introduced in this work, the Profile Rotator (PR), which rigidly rotates the beam profile (including its polarization structure) using an appropriately calibrated K-mirror and a Faraday rotator. The theoretical framework is based on the analysis of mode symmetries and their interaction with cavity boundary conditions. This work lays the groundwork for experimental implementations of efficient mode sorting, with potential applications in high-dimensional quantum information processing, optical communications, and structured light manipulation.

\end{abstract}
\pacs{}
\maketitle
%
\section{Introduction}

The manipulation and control of structured light have opened new avenues for both classical and quantum optical technologies. Among structured light fields, vector vortex beams (VVB) stand out due to their spatially varying polarization \cite{Rosales2018,Zhan2009}, offering a rich platform for encoding information \cite{Ndagano2018,Borges2010,Pereira2014,Dambrosio2012,Ambrosio2016} and probing fundamental physical phenomena \cite{Abouraddy2006,Roxworthy2010,Fatemi2011,Neugebauer2014,Parigi2015}.

The advantage of using VVB encoding lies in the nonseparable coupling between spatial and polarization degrees of freedom \cite{Milione2015b}. Unlike scalar orbital angular momentum (OAM) modes, which are separable in these variables, VVBs exhibit spatially varying polarization structures that can be tailored to exploit specific symmetries or interactions in optical systems. This coupling enables novel forms of mode selectivity, robustness against certain perturbations\cite{yuan2022, Cheng2023, cheng2009}, and access to transformations that act jointly on polarization and spatial components \cite{Rubano2019}, which are not achievable with scalar OAM modes alone.

Efficiently sorting and distinguishing VVB modes is a key step for applications ranging from high-dimensional quantum communication to advanced imaging techniques. In this context, developing robust methods to separate vector optical modes based on their intrinsic properties, such as symmetry, is particularly important.

Several methods have been proposed for sorting vector vortex beams (VVBs), though the literature on the subject remains limited. One approach, reported in 2015 \cite{Milione2015}, employs a series of Sagnac interferometers combined with q-plates to sort VVBs according to their topological charge $m$. In this scheme, that sorts $N$ distinct modes with $N - 1$ cascaded Sagnac interferometers, the spatial structure of the beams is preserved after separation, but, by design, separation of two modes of the same order is not achieved. A different method, introduced in 2019 \cite{Jia2019}, uses a simpler architecture based on one beam splitter, two q-plates \cite{Rubano2019}, two single-mode fibers, and two polarizing beam splitters to sort the four elements of the first-order VVB basis. However, this scheme inherently discards 50\% of the optical power and yields Gaussian outputs rather than preserving the vectorial nature of the input beams.

More recently, our research group has experimentally demonstrated that a triangular optical cavity functions as an effective parity sorter, capable of distinguishing structured light modes (including vector vortex beams) based on their symmetry properties \cite{Santos2021,Rodrigues2024,Fagundes2025}. In particular, we have verified that vector vortex beams are resonant with triangular cavities, opening new possibilities for sorting and manipulating these complex beams in a compact, power-preserving, and mode-preserving manner.

In the present work, we propose an experiment using three optical cavities to sort a collection of four vector vortex modes of the same arbitrary order $m$. The proposed setup begins with a primary cavity that separates the modes into two pairs: one pair symmetric and the other antisymmetric with respect to a horizontal flip. The symmetric pair is transmitted, while the antisymmetric pair is reflected. Each output arm then feeds into a secondary cavity, which further sorts the pairs into individual modes based on their parity. By combining the symmetry properties of the vector modes with the parity-selective nature of cavity reflections and transmissions, our method enables a complete separation of the four input modes.

A crucial aspect of our approach involves the detailed study of the symmetries of vector vortex modes with respect to radial axes at varying angles. This analysis guides the optimal alignment of the cavities and the necessary polarization transformations to ensure efficient mode sorting.

This proposal lays the theoretical groundwork for an experimentally feasible mode-sorting scheme. In Section \ref{sec:VVB}, we define the first-order VVB basis and investigate the parity properties of its four elements. In Section \ref{sec:exp}, an experiment is designed to explore these properties and separate any superposition of first-order VVBs. Section \ref{sec:higher-order} generalizes Section \ref{sec:VVB} to arbitrary order $m$ and shows that the same experimental scheme can be used to sort any superposition of VVBs of the same order $m$.

\section{Parity of First-Order Vector Vortex Beams}
\label{sec:VVB}

A first-order VVB basis can be defined in terms of the first-order Laguerre-Gauss beams $LG_{\pm10}$ (azimuthal phase $\e^{\mp\ii \varphi}$ and radial number 0) with circular polarization \cite{Maurer2007}:
\begin{align}
    \Psi^+_1&=LG_{10}\,\hat{\text{e}}_R\,+\,LG_{-10}\,\hat{\text{e}}_L\,, \label{eq:Psi+}\\
    \Psi^-_1&=LG_{10}\,\hat{\text{e}}_R\,-\,LG_{-10}\,\hat{\text{e}}_L\,, \label{eq:Psi-}\\
    \Phi^+_1&=-\left(LG_{-10}\,\hat{\text{e}}_R\,+\,LG_{10}\,\hat{\text{e}}_L\right)\,, \label{eq:Phi+}\\
    \Phi^-_1&=-\left(LG_{-10}\,\hat{\text{e}}_R\,-\,LG_{10}\,\hat{\text{e}}_L\right)\,, \label{eq:Phi-}
\end{align}
where $\hat{\text{e}}_R$ and $\hat{\text{e}}_L$ are the unit vectors for right ($R$) and left ($L$) circular polarizations. 

\begin{figure}[h!]
    \includegraphics[width=\columnwidth]{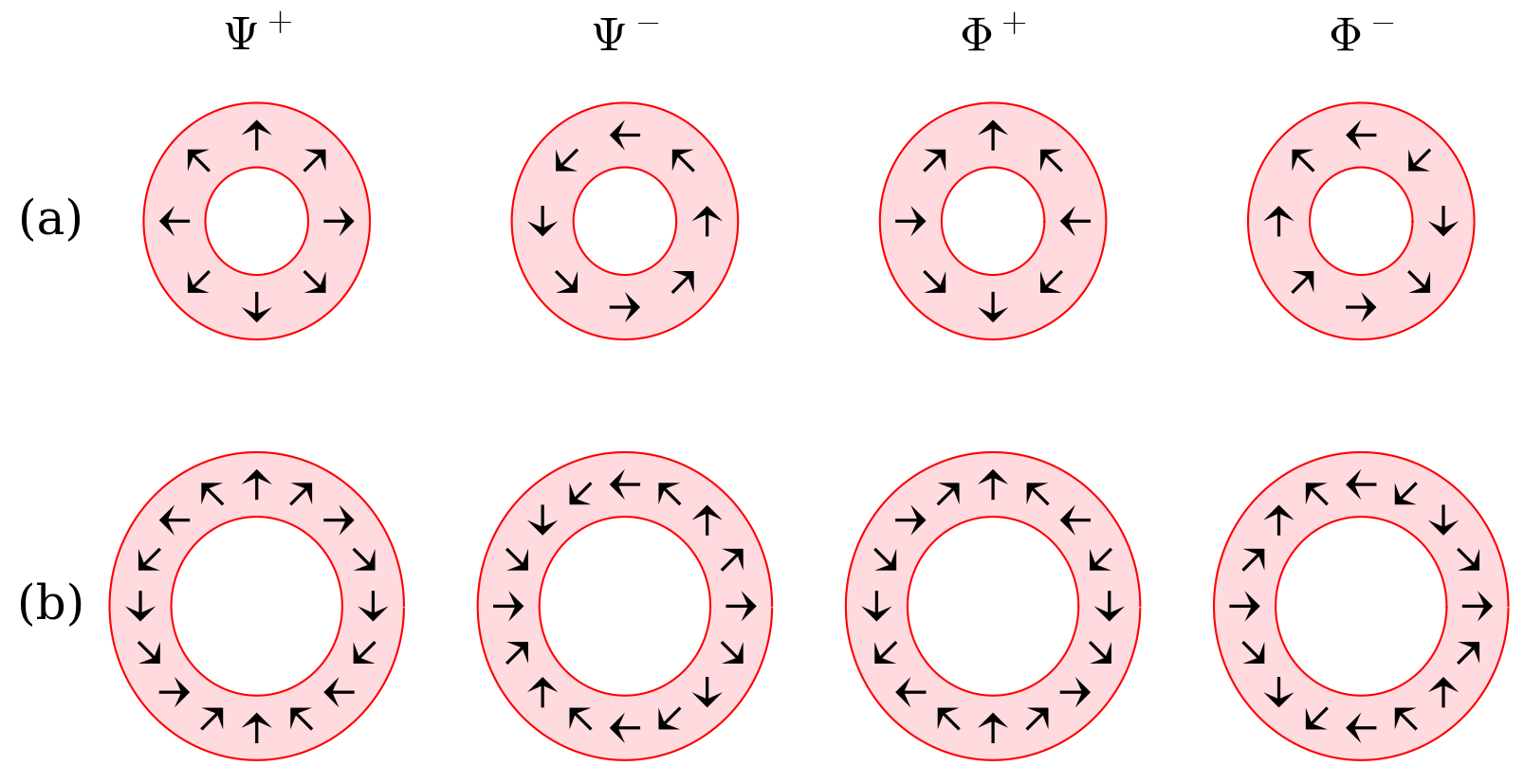}
    \caption{Representation of the ring-shaped intensity profile and inhomogeneous polarization of Vector Vortex Beam bases for (a) $m=1$ and (b) $m=2$. From left to right: $\Psi^+_m$, $\Psi^-_m$, $\Phi^+_m$, and $\Phi^-_m$.}
    \label{fig:vvb}
\end{figure}

The four modes defined above exhibit the intensity profile of an $LG_{10}$ mode, which is a ring-shaped pattern; nevertheless, their polarizations are not homogeneous and depend solely on the azimuthal angle $\varphi$, defined here with respect to the vertical direction. Fig. \ref{fig:vvb}a illustrates the resulting inhomogeneous polarization over the beam profiles of modes $\Psi^+_1$ (radial), $\Psi^-_1$ (azimuthal), $\Phi^+_1$ and $\Phi^-_1$.

Let us describe the symmetry properties of each element of this basis. The symmetry we are interested in is parity symmetry with respect to a horizontal flip. A horizontal flip consists on changing the sign of horizontal unit vector of the reference frame ($\hat{\e}_x\rightarrow -\hat{\e}_x$), effectively mirroring the electric field's transverse profile about the vertical axis. This operation physically relates to a beam reflecting with non-normal incidence on a plane mirror parallel to a vertical plane and is key to understand why a triangular cavity can be thought of as a parity sorter \cite{Santos2021,Rodrigues2024,Fagundes2025}.

Modes $\Psi^+_1$ and $\Phi^+_1$ are, each, mirror images of themselves, so we say they have even parity symmetry with respect to the vertical axis. Mathematically, this can be expressed as $\mathcal{P}(\Psi^+_1)=+\Psi^+_1$ and $\mathcal{P}(\Phi^+_1)=+\Phi^+_1$, where $\mathcal{P}$ is the parity operator. In contrast, applying a horizontal flip to modes $\Psi^-_1$ and $\Phi^-_1$ results in reversing the sense of the electric field vector at every single point in the profile, while maintaining its direction. For that reason, we say theses modes exhibit odd parity symmetry (or antisymmetry), and we write $\mathcal{P}(\Psi^-_1)=-\Psi^-_1$, $\mathcal{P}(\Phi^-_1)=-\Phi^-_1$.

It is important to note that changing the reference axis used to analyze symmetry can alter the parity of a mode. For instance, while $\Phi^+_1$ is symmetric about the axis $\varphi=0$ (vertical), it is antisymmetric with respect to axis $\varphi=\pi/4$ (antidiagonal), which can be expressed as $\mathcal{P}_{\pi/4}(\Phi^+_1)=-\Phi^+_1$. Other angles may provide axes with respect to which modes lose parity.

By a simple visual inspection, it is possible to find all symmetry/antisymmetry axes of all four modes. A rigorous demonstration is left for when we generalize this result for order $m$ in Section \ref{sec:higher-order}. Table \ref{tab:symmetry-axes-1} summarizes the results.


\begin{table}
\renewcommand{\arraystretch}{1.5}
\begin{tabularx}{\columnwidth}{ >{\centering}X  >{\centering}X  >{\centering}X >{\centering}X  >{\centering\arraybackslash}X}
           & \multicolumn{2}{c}{\cellcolor{gray!30} Symmetry axes}  & \multicolumn{2}{c}{\cellcolor{gray!30} Antisymmetry axes} \\ \hline
Mode       & Angles $\alpha$  & Total \# & Angles $\alpha$    & Total \#  \\ \hline
$\Psi^+_1$ & $[0,\,\pi[$      & $\infty$ & $\emptyset$        & 0         \\
$\Psi^-_1$ & $\emptyset$      & 0        & $[0,\,\pi[$        & $\infty$  \\
$\Phi^+_1$ & $0,\ \pi/2$      & 2        & $\pi/4,\ 3\pi/4$   & 2         \\
$\Phi^-_1$ & $\pi/4,\ 3\pi/4$ & 2        & $0,\ \pi/2$        & 2         \\ \hline
\end{tabularx}
\caption{Symmetry and Antisymmetry axes for the VVB modes of the first order $(m=1)$. ``Angles $\alpha$'' columns indicate the values of axes angles $\alpha$ with respect to vertical. ``Total \#'' is the total number of symmetry or antisymmetry axes.}
\label{tab:symmetry-axes-1}
\end{table}


\section{A proposed experiment}
\label{sec:exp}

In this section, we propose an experiment that explores the parity properties of first-order VVBs to sort them out in a non destructive way, given any arbitrary linear combination of these four basis elements. This proposal is based on results previously obtained by our own group: first-order VVBs can resonate in a triangular optical cavity \cite{Rodrigues2024} and, more importantly, triangular cavities are parity sorters for spatial modes of the same order \cite{Santos2021, Rodrigues2024, Fagundes2025}. 

\begin{figure}[H]
   \includegraphics[width=\columnwidth]{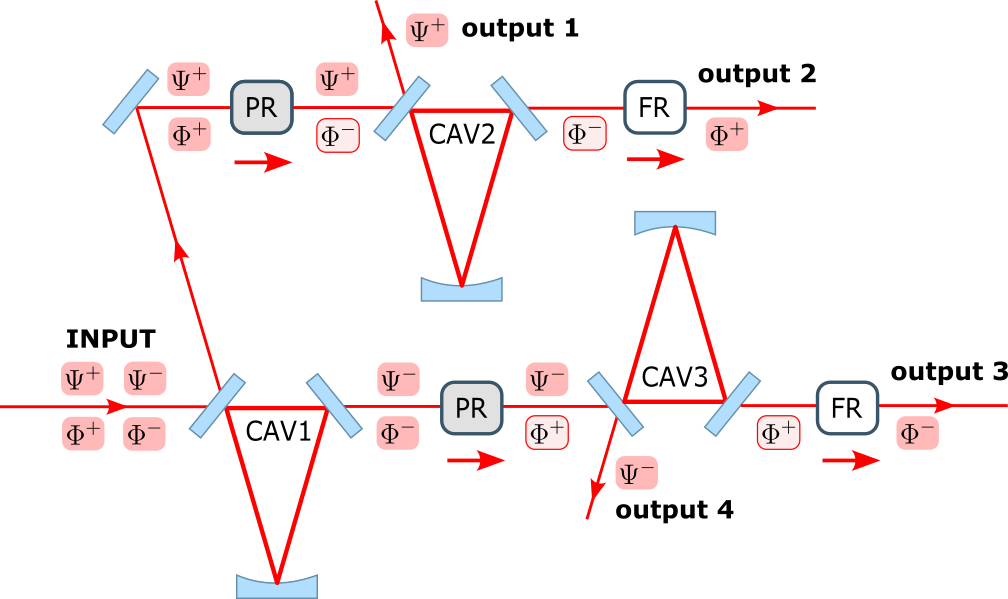}
   \caption{Proposed experimental setup. Any linear combination of the four elements of the vector mode basis can be sorted by a set of three cavities. Cavity-resonant modes are fully transmitted and non-resonant modes are fully reflected, provided the plane cavity mirrors have the same reflectivity and the concave mirror is highly reflective. PR = $\pi/4$ profile rotator: rigidly rotates the beam profile (including polarization pattern) by an angle of $\pi/4$ before cavities 2 and 3. FR = $\pi/2$ Faraday rotator: locally rotates the polarization vectors by $\pi/2$ all over the beam profile.}
   \label{fig:setup}
\end{figure}

The proposed experiment, illustrated in Fig. \ref{fig:setup}, involves the use of three triangular cavities and two steps. In the first step, a single cavity (CAV1) separates the four modes into two pairs: the modes with even parity ($\Psi^+_1,\ \Phi^+_1$) are, say, transmitted, while those with odd parity ($\Psi^-_1,\ \Phi^-_1$) are reflected. In the second step, two other cavities, CAV2 and CAV3 (aligned with the reflection and the transmission of CAV1, respectively) are used to split each pair of modes into two separate modes. The result is that the four modes, initially propagating within the same beam, are ultimately demultiplexed and propagate independently in different directions.

In order to separate modes with the same parity, say $\Psi^+_1$ and $\Phi^+_1$, an additional mechanism called profile rotator (PR), discussed in the next Section, is needed before CAV2 that rigidly rotates the beam profile about the optical axis by an angle of $\pi/4$. This effectively transforms mode $\Phi^+_1$ into $\Phi^-_1$, while preserving mode $\Psi^+_1$. Now, the two copropagating modes have opposite parity and can be separated by CAV2. An additional profile rotator (at $-\pi/4$) may be placed at the right output of the cavity to restore mode $\Phi^+_1$. The same idea applies to the pair $\Psi^-_1$ and $\Phi^-_1$: although both have odd parity, a profile rotator at $\pi/4$ can operate on them so that $(\Psi^-_1,\Phi^-_1)\rightarrow (\Psi^-_1,\Phi^+_1)$, and the cavity splitting becomes possible on CAV3.

By construction, the reference axis is defined as the one perpendicular to the cavity plane. Typically, the cavity plane is horizontal, making the reference axis vertical. The cavity then always sorts symmetric and antisymmetric modes relative to this vertical axis. In other words, mode sorting succeeds when the reference axis aligns with the symmetry/antisymmetry axis of the modes. After the first cavity, this is precisely the role of the profile rotator: it brings the symmetry/antisymmetry axis from the diagonal into the vertical direction. An alternative would be to rotate the cavity plane itself by 45$^\circ$, but this would introduce the complication of having the reflected (non-resonant) beam completely off the horizontal plane.

Making this three-cavity mode sorter fully functional involves actively keeping the cavities in resonance. The resonance lengths of the symmetric and antisymmetric modes are regularly interleaved, and the spacing between adjacent resonance peaks is half a wavelength. If the cavity is locked to a resonance peak of the symmetric mode, this mode will be transmitted and, consequently, the antisymmetric mode will be reflected. However, nothing prevents the user from locking the cavity to the resonance of the antisymmetric mode, in which case it is transmitted and the symmetric mode is reflected. It is, therefore, a matter of choice.

Locking can be achieved using a simple dither-locking technique, in which the position of one cavity mirror is modulated by a piezoelectric actuator. This modulation produces an error signal after demodulation of a leakage signal (transmitted by the concave mirror, for instance). The dither-locking method operates at a low modulation frequency (typically in the kHz range, as opposed to the MHz range used in the more complex Pound-Drever-Hall technique) and is well suited for low-finesse cavities. The alignment and mode matching of the three cavities can be performed and optimized using a zero-order Gaussian beam with the same waist $w_0$ as that of the vector beams. An impedance-matched, low-finesse cavity with a highly reflective concave mirror is key to maximize transmittance, and therefore, mode separation efficiency.

\section{Profile Rotator}
\label{sec:PR}

Profile rotation (PR) is required in our mode sorting scheme to break the symmetry of the mode pairs before CAV2 and CAV3. A PR of $\pi/4$ effectively transforms the pair of symmetric modes before CAV2 (or antisymmetric modes before CAV3) into one symmetric and one antisymmetric mode, enabling mode separation. 

Recently, McWilliam \textit{et al.} have shown it is possible to implement a profile rotator using a set of four mirrors \cite{McWilliam2022} at the expense of causing a lateral translation and a change in the height of the beam. In this section, we propose and discuss the implementation of a PR composed of a K-mirror and a Faraday rotator that manages to rotate the beam while keeping the original optical axis.

K-mirrors are optical elements composed of three plane mirrors disposed in a K-shape and are commonly employed to rotate the transverse beam profile by a desired angle about the optical axis, while maintaining beam alignment. The rotation angle is controlled by rotation of the optical element itself: the transverse profile rotates by an angle $2\beta$ when the optical element is rotated by $\beta$. In usual applications, input polarization is homogeneous over the beam profile and stays homogeneous at the output, although it may change due to reflections at non-normal incidence. 

It has been shown in \cite{Karan2022,Karan2024} that the polarization change induced by a K-mirror can be minimized by appropriately selecting the mirror angles and the type of metal coating. For example, at the wavelength of 780~nm used in our lab, we have calculated -- using the method proposed in \cite{Karan2022} -- that with silver-coated mirrors and an angle of 13.56$^\circ$ of the first and third mirrors relative to the optical axis, the mean polarization change can be reduced to as little as $0.003\pi$ \cite{code_Kmirror_2025}, while still achieving the desired profile rotation. This small base angle implies a relatively long, yet practical, K-mirror system (approximately 15 cm in length). The required length may vary depending on the operating wavelength and mirror coating material, and users can take advantage of modern 3D-printing technologies to design and fabricate custom K-mirror mounts tailored to their specific requirements.

Applying the transformation of a polarization maintaining K-mirror to vector beam brings a subtlety: because of the inhomogenous polarization over the transverse profile, the polarization at a given point A is brought to point B without rotation of the polarization vector, changing the overall polarization distribution. More precisely, if the K-mirror performs a $2\beta$-rotation on the wavefront of a beam with polarization $\theta(\varphi)$, the output polarization distribution reads $\theta'(\varphi)=\theta(\varphi-2\beta)$.

\begin{figure}[h!]
    \includegraphics[width=\columnwidth]{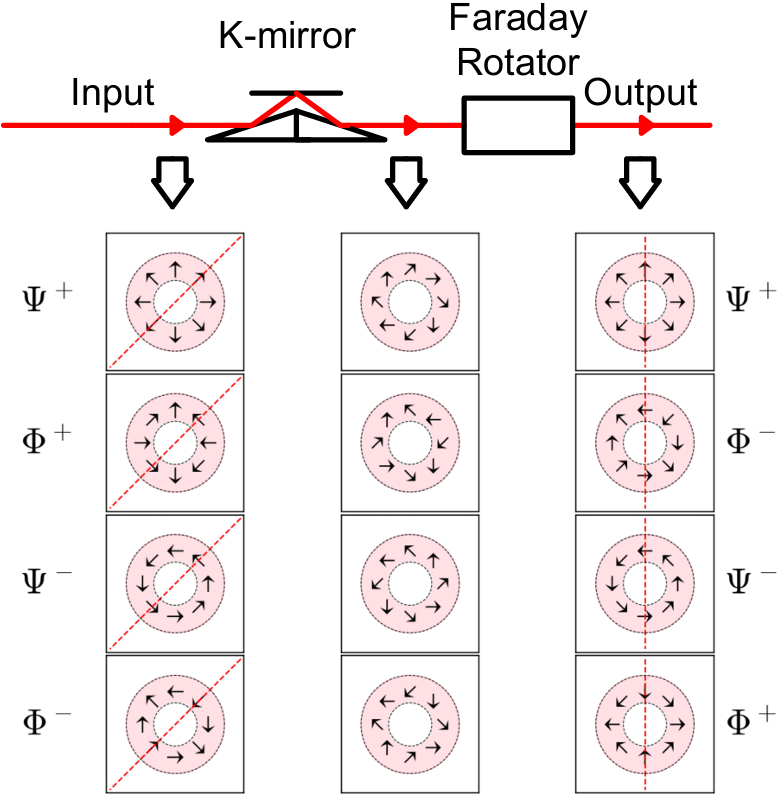}
    \caption{Schematic of the proposed Profile Rotator (PR), showing its effect on the polarization profile of $m = 1$ vector vortex beams after each component. The K-mirror is set to a rotation angle $\beta = \pi/8$, and the Faraday rotator applies a uniform polarization rotation of $\pi/4$ radians. All angles are measured counterclockwise from the vertical.}
    \label{fig:BeamRotation}
\end{figure}

To achieve full rigid profile rotation (PR), we propose using a combination of a K-mirror set at a rotation angle of $\beta = \pi/8$, together with a Faraday rotator that uniformly rotates the polarization by $\pi/4$ radians across the entire beam. Figure~\ref{fig:BeamRotation} illustrates the resulting polarization states after the beam passes through each optical element, for all four vector vortex beam (VVB) basis states with $m = 1$ as inputs.




\section{Extension to higher orders}
\label{sec:higher-order}

In this section, we define higher-order VVB modes and find their symmetry properties. With this, we study the viability of extending the mode sorter presented above to the case of arbitrary order $m$.

\subsection{Arbitrary-Order Vector Vortex Beams}
\label{sec:VVB-m}

For each given order $m$, it is possible to construct an orthogonal basis composed of exactly four VVB's $\{\Psi^\pm_m,\Phi^\pm_m\}$ that generalizes $m=1$ basis shown in Fig. \ref{fig:vvb}. The basis elements write as follows \cite{Milione2011}:
\begin{align}
    \Psi^+_m&=LG_{m0}\,\hat{\text{e}}_R\,+\,LG_{-m0}\,\hat{\text{e}}_L\,, \label{eq:Psi+m}\\
    \Psi^-_m&=LG_{m0}\,\hat{\text{e}}_R\,-\,LG_{-m0}\,\hat{\text{e}}_L\,, \label{eq:Psi-m}\\
    \Phi^+_m&=-\left(LG_{-m0}\,\hat{\text{e}}_R\,+\,LG_{m0}\,\hat{\text{e}}_L\right)\,, \label{eq:Phi+m}\\
    \Phi^-_m&=-\left(LG_{-m0}\,\hat{\text{e}}_R\,-\,LG_{m0}\,\hat{\text{e}}_L\right)\,, \label{eq:Phi-m}
\end{align}
where $LG_{\pm m0}$ represents a Laguerre-Gauss mode with azimuthal phase $\e^{\mp \ii m\varphi}$ and radial number and 0.

As in the $m=1$ case, the four modes defined above exhibit the intensity profile of an $LG_{m0}$ mode, a ring-shaped pattern whose radius increases as $\sqrt{m}$, and their polarizations only depend on the azimuthal angle $\varphi$ (see Fig. \ref{fig:vvb}b for depiction of $m=2$ basis). From the equations above, it is easy to show that the polarization angle $\theta(\varphi)$ with respect to vertical direction, for each of these modes is
\begin{align}
    \Psi^+_m:&\ \theta(\varphi)=m\varphi\,, \label{eq:ThetaPsi+}\\
    \Psi^-_m:&\ \theta(\varphi)=m\varphi+\pi/2\,, \label{eq:ThetaPsi-}\\
    \Phi^+_m:&\ \theta(\varphi)=-m\varphi\,, \label{eq:ThetaPhi+}\\
    \Phi^-_m:&\ \theta(\varphi)=-m\varphi+\pi/2\,. \label{eq:ThetaPhi-}
\end{align}

\subsection{Symmetry Properties of Vector Vortex Beams}
\label{sec:symmetry-m}

As in the $m=1$ case, modes $\Psi^+_m$ and $\Phi^+_m$ are, each, mirror images of themselves, so they have even parity symmetry with respect to the vertical axis. In contrast, applying a horizontal flip to modes $\Psi^-_m$ and $\Phi^-_m$ results in reversing the sense of the electric field vector at every single point in the profile, while maintaining its direction: they are antisymmetric.

Let us rigorously demonstrated this. Since the vector modes considered here all have cylindrically symmetric intensity profiles, it is sufficient to analyze their parity by looking solely at the polarization angle $\theta(\varphi)$. In fact, such a vector vortex mode is symmetric/antisymmetric with respect to the vertical axis if, and only if, 
\begin{align}
    \theta(\varphi)+\theta(-\varphi)&=2k\pi\,, &\ \textbf{[sym.]}\label{eq:symmetry}\\
    \theta(\varphi)+\theta(-\varphi)&=\pi+2k\pi\,, &\ \textbf{[antisym.]}\label{eq:antisymmetry}
\end{align}
with $k\in\mathbb{Z}$. It follows directly from eqs. (\ref{eq:ThetaPsi+}) to (\ref{eq:ThetaPhi-}) that modes $\Psi^+_m$ and $\Phi^+_m$ are symmetric, while $\Psi^-_m$ and $\Phi^-_m$ are antisymmetric, $\forall  m$.

Let us now find all possible symmetry/antisymmetry axes for each mode (and arbitrary $m$). If an axis forms an angle $\alpha$ with the vertical, we call it a symmetry axis of a given mode $M$ if $\mathcal{P}_\alpha(M)= M$, where $\mathcal{P}_\alpha$ is the parity operator with respect to that axis. Similarly, we call this axis an antisymmetry axis for mode $M$ if $\mathcal{P}_\alpha(M)= -M$. In terms of the polarization angles, the conditions for angle $\alpha$ to define a symmetry/antisymmetry axis are obtained by generalizing eqs. (\ref{eq:symmetry}) and (\ref{eq:antisymmetry}):
\begin{align}
    \theta(\alpha+\varphi)+\theta(\alpha-\varphi)-2\alpha&=2k\pi\,, \label{eq:symmetry-alpha}\\
    \theta(\alpha+\varphi)+\theta(\alpha-\varphi)-2\alpha&=\pi+2k\pi\,. \label{eq:antisymmetry-alpha}
\end{align}
Finding all possible symmetry and antisymmetry axes of a given mode consists of solving, separately, the above equations for $\alpha$ (mod $\pi$).

For instance, in order to find all symmetry axis of mode $\Psi^+_m$, one substitutes eq. (\ref{eq:ThetaPsi+}) in eq. (\ref{eq:symmetry-alpha}), which gives $(m-1)\alpha=k\pi$, $k\in\mathbb{Z}$. For $m=1$, this equation only holds if $k=0$, in which case, any $\alpha$ is a solution, as expected for the radially symmetric beam $\Psi^+_1$ (see Fig. \ref{fig:vvb}a). For $m\geq 2$, we find $\alpha=k\pi/(m-1)$, meaning that there are $m-1$ symmetry axes for $\Psi^+_{m\geq 2}$ ($k=0,\ 1,\ \dots,\ m-2$, since it is redundant to take values of $k$ from $m-1$ on). As an example, take $m=4$: there are 3 distinct symmetry axes, namely, $\alpha=0,\ \pi/3,\ 2\pi/3$. 

\begin{figure}[h!]
    \includegraphics[width=\columnwidth]{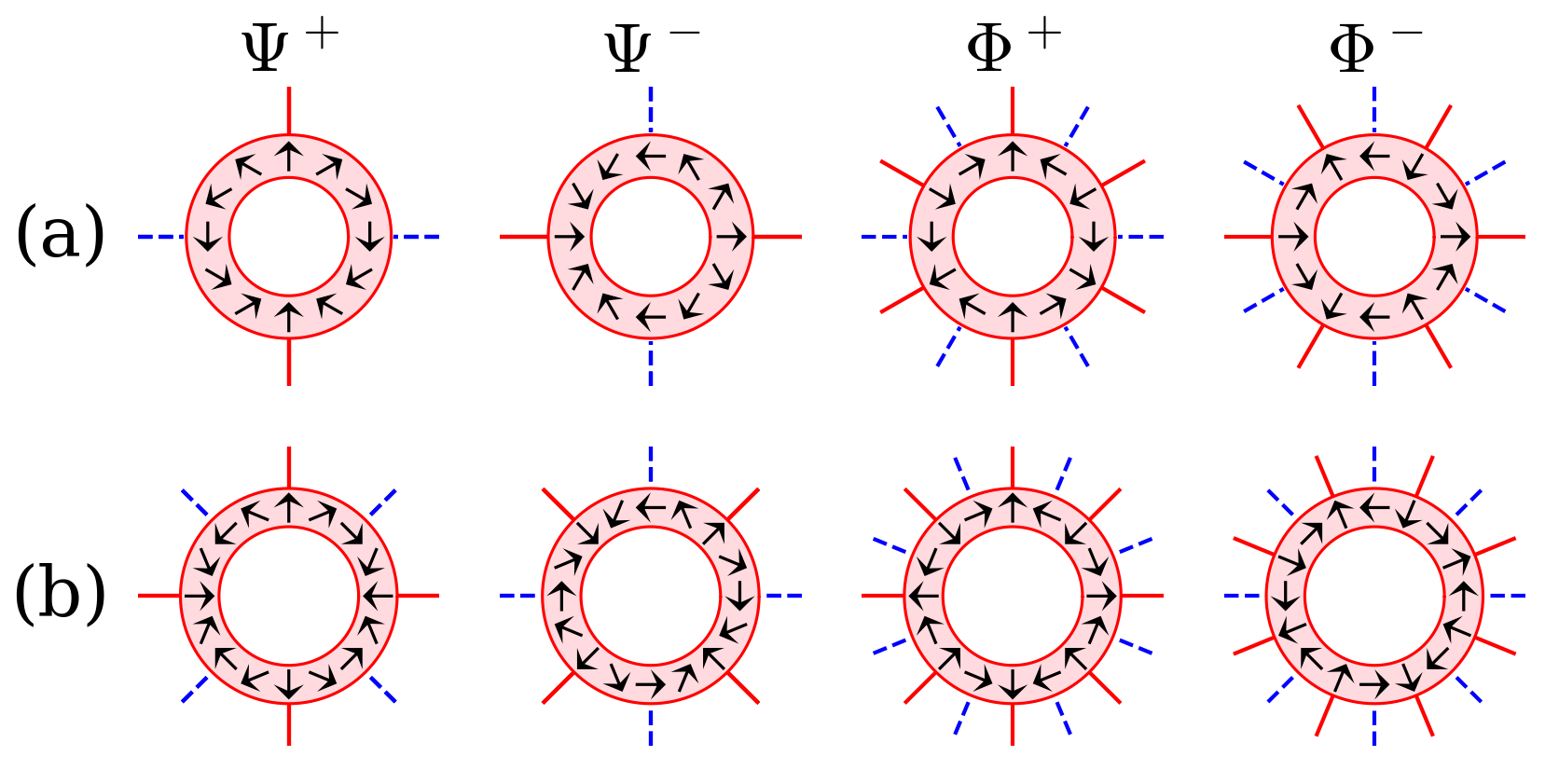}
    \caption{Representation of the symmetry (red lines) and antisymmetry (blue dashed lines) axes for (a) $m=2$ and (b) $m=3$ VVBs. The axes are described by the angle $\alpha$ they make with the vertical direction, measured counterclockwise.}
    \label{fig:symmetry-axes}
\end{figure}

Similar calculations for modes $\Psi^-_m$, $\Phi^+_m$ and $\Phi^-_m$, using eqs. (\ref{eq:ThetaPsi-}), (\ref{eq:ThetaPhi+}) and (\ref{eq:ThetaPhi-}), respectively, also give the number of symmetry axes and their angles $\alpha$ with respect to the vertical. The antisymmetry axes may also be calculated for each mode: it suffices to replace their respective equation -- from (\ref{eq:ThetaPsi+}) to (\ref{eq:ThetaPhi-}) -- in eq. (\ref{eq:antisymmetry-alpha}) and find the possible values of $\alpha$ obtained from varying the integer $k$. The results are summarized in Table~\ref{tab:symmetry-axes-m} and graphically exemplified in Fig.~\ref{fig:symmetry-axes}.

\begin{table}
\renewcommand{\arraystretch}{1.5}
\begin{tabularx}{\columnwidth}{ >{\centering}X  >{\centering}X  >{\centering}X >{\centering}X  >{\centering\arraybackslash}X}
           & \multicolumn{2}{c}{\cellcolor{gray!30} Symmetry axes}  & \multicolumn{2}{c}{\cellcolor{gray!30} Antisymmetry axes} \\ \hline
Mode       & Angles $\alpha$           & Total \# & Angles $\alpha$           & Total \#  \\ \hline
\noalign{\vskip 1ex}
$\Psi^+_m$ & $\dfrac{k\pi}{m-1}$        & $m-1$    & $\dfrac{(k+1/2)\pi}{m-1}$  & $m-1$   \\ [2ex]
$\Psi^-_m$ & $\dfrac{(k+1/2)\pi}{m-1}$  & $m-1$    & $\dfrac{k\pi}{m-1}$        & $m-1$   \\ [2ex]
$\Phi^+_m$ & $\dfrac{k\pi}{m+1}$        & $m+1$    & $\dfrac{(k+1/2)\pi}{m+1}$  & $m+1$   \\ [2ex]
$\Phi^-_m$ & $\dfrac{(k+1/2)\pi}{m+1}$  & $m+1$    & $\dfrac{k\pi}{m+1}$        & $m+1$   \\ [2ex] \hline
\end{tabularx}
\caption{Symmetry and Antisymmetry axes for all VVB modes of higher order $(m\geq2)$. ``Angles $\alpha$'' columns indicate the values of axes angles $\alpha$ with respect to vertical. ``Total \#'' is the total number of symmetry or antisymmetry axes.}
\label{tab:symmetry-axes-m}
\end{table}

\subsection{A generalized vector mode sorter?}
\label{sec:mode-sorter-m}

A generalized vector mode sorter would consist of an optical system capable of demultiplexing the four elements of the vector mode basis for any given $m$. Such a sorter must satisfy two basic requirements:
\begin{enumerate}
    \item Each mode $\Psi^+_m$, $\Psi^-_m$, $\Phi^+_m$, and $\Phi^-_m$ must fully resonate with the cavity at some length, in order to preserve its profile upon transmission;
    \item There must be a rotation angle $\alpha$ for which parity can be split, i.e. modes that share the same parity with respect to the vertical exhibit opposite parity with respect to the axis oriented at angle $\alpha$.
\end{enumerate}

The \textbf{first requirement} is fulfilled for every $m$. In reference \cite{Rodrigues2024}, we show that this is the case for $m=1$. Indeed, each vector mode $\Psi^+_1$, $\Psi^-_1$, $\Phi^+_1$, or $\Phi^-_1$ may be written as a sum of their horizontal and vertical polarization components. For a given mode, both components are Hermite-Gaussian modes having order $1$ and the same parity. For instance, the radial mode writes $\Psi^+_1\propto HG_{10}\,\hat{\e}_x+HG_{01}\,\hat{\e}_y$. Sharing the same order implies that both components accumulate the same Gouy phase \cite{Gouy1890,Kogelnik1966} per cavity round trip. Having the same parity ensures that the phase accumulated per round trip due to reflections is also identical for both components. As a result, both polarization components resonate simultaneously, making the full vector mode resonant with the cavity. This reasoning extends to higher-order modes: for all $m\geq 2$, resonance is preserved because the horizontal and vertical components continue to share both the same order and parity.

Let us now investigate whether the \textbf{second requirement} is satisfied. In the proposed experiment, angle $\alpha=\pi/4$ is suggested to realize the parity-splitting rotation for $m=1$. It turns out that this angle also splits parity for every \textbf{odd} $m$. Using the results presented in Table \ref{tab:symmetry-axes-m}, it is possible to verify that, 
\begin{itemize}
    \item if $m=4n+1$, then $\alpha=\pi/4$ defines a symmetry axis for $\Psi^+_m$ and $\Phi^-_m$ and an antisymmetry axis for $\Phi^+_m$ and $\Psi^-_m$; while
    \item if $m=4n-1$, then $\alpha=\pi/4$ defines an antisymmetry axis for $\Psi^+_m$ and $\Phi^-_m$ and a symmetry axis for $\Phi^+_m$ and $\Psi^-_m$,
\end{itemize}
where $n\in\mathbb{N}^*$.

Consequently, a $\pi/4$ beam profile rotation performs the transformation 
$$(\Psi^+_m,\ \Phi^+_m,\ \Psi^-_m,\ \Phi^-_m)\rightarrow (\Psi^+_m,\ \Phi^-_m,\ \Psi^-_m,\ \Phi^+_m)$$
for $m=4n+1$ and
$$(\Psi^+_m,\ \Phi^+_m,\ \Psi^-_m,\ \Phi^-_m)\rightarrow (\Psi^-_m,\ \Phi^+_m,\ \Psi^+_m,\ \Phi^-_m)$$
for $m=4n-1$, up to a $\pi$ phase shift applied to certain modes depending on their order, which does not affect the performance of the mode sorter, since it is insensitive to phase.

Above, we argued that a beam profile rotation of $\alpha = \pi/4$ achieves the desired parity splitting for any \textbf{odd} $m$, enabling mode separation at cavities 2 and 3. However, for \textbf{even} $m$, no angle of beam profile rotation yields such parity splitting. As seen from the symmetry and antisymmetry conditions in Table \ref{tab:symmetry-axes-m}, it is impossible to find a value of $\alpha$ that satisfies the cavity splitting requirements for even-order modes. Nevertheless, as shown in \cite{Jia2019}, even-order modes can be converted to the next higher (odd) order using a single q-plate with $q = 1/2$. Thus, for even-order modes, it suffices to insert such a q-plate before the mode sorter to enable separation, followed by another q-plate with $q = -1/2$ after the sorter to restore the original order. In this way, the mode sorting scheme can be extended to operate for arbitrary values of $m$.

\section{Conclusion}
\label{sec:conc}

In summary, we have proposed an experimental scheme for separating vector vortex beams (VVBs) of arbitrary order $m$. We show that any superposition of the four elements of the order-$m$ VVB basis can be transformed into four independently propagating beams that preserve their original polarization characteristics.

This separation is enabled by a combination of three resonant triangular cavities -- sensitive to the spatial symmetry of the modes -- and a novel system introduced in this work, the Profile Rotator (PR). The PR, composed of a K-mirror and a Faraday rotator, performs a rigid rotation of the beam profile, including its polarization pattern. To the best of our knowledge, this is the first time K-mirrors have been employed in the context of vector vortex beams.

Beyond its fundamental interest, our proposed method is introduced as a new tool that can be used either on its own or in combination with existing techniques for practical applications. While the setup proposed can only sort four modes it could be used for communication protocols, as shown in \cite{Milione2015}. Should higher dimensional applications prove necessary, it could be combined with other methods, as the one in \cite{Jia2019}, to first sort VVBs based on their order and then into the individual modes.

By establishing a systematic framework for mode separation based on symmetry and cavity interactions, this work bridges theoretical insight and experimental feasibility, opening new avenues for the control and manipulation of structured light.

\begin{acknowledgments}
We acknowledge funding from the Brazilian agencies: Conselho Nacional de Desenvolvimento Tecnol\'ogico (CNPq), Coordena\c c\~{a}o de Aperfei\c coamento de Pessoal de N\'ivel Superior - Brasil (CAPES) and Instituto Nacional de Ci\^encia e Tecnologia de Informa\c c\~ao Qu\^antica (INCT/IQ 465469/2014-0).
\end{acknowledgments}


\begin{thebibliography}{30}%
\makeatletter
\providecommand \@ifxundefined [1]{%
 \@ifx{#1\undefined}
}%
\providecommand \@ifnum [1]{%
 \ifnum #1\expandafter \@firstoftwo
 \else \expandafter \@secondoftwo
 \fi
}%
\providecommand \@ifx [1]{%
 \ifx #1\expandafter \@firstoftwo
 \else \expandafter \@secondoftwo
 \fi
}%
\providecommand \natexlab [1]{#1}%
\providecommand \enquote  [1]{``#1''}%
\providecommand \bibnamefont  [1]{#1}%
\providecommand \bibfnamefont [1]{#1}%
\providecommand \citenamefont [1]{#1}%
\providecommand \href@noop [0]{\@secondoftwo}%
\providecommand \href [0]{\begingroup \@sanitize@url \@href}%
\providecommand \@href[1]{\@@startlink{#1}\@@href}%
\providecommand \@@href[1]{\endgroup#1\@@endlink}%
\providecommand \@sanitize@url [0]{\catcode `\\12\catcode `\$12\catcode `\&12\catcode `\#12\catcode `\^12\catcode `\_12\catcode `\%12\relax}%
\providecommand \@@startlink[1]{}%
\providecommand \@@endlink[0]{}%
\providecommand \url  [0]{\begingroup\@sanitize@url \@url }%
\providecommand \@url [1]{\endgroup\@href {#1}{\urlprefix }}%
\providecommand \urlprefix  [0]{URL }%
\providecommand \Eprint [0]{\href }%
\providecommand \doibase [0]{https://doi.org/}%
\providecommand \selectlanguage [0]{\@gobble}%
\providecommand \bibinfo  [0]{\@secondoftwo}%
\providecommand \bibfield  [0]{\@secondoftwo}%
\providecommand \translation [1]{[#1]}%
\providecommand \BibitemOpen [0]{}%
\providecommand \bibitemStop [0]{}%
\providecommand \bibitemNoStop [0]{.\EOS\space}%
\providecommand \EOS [0]{\spacefactor3000\relax}%
\providecommand \BibitemShut  [1]{\csname bibitem#1\endcsname}%
\let\auto@bib@innerbib\@empty
\bibitem [{\citenamefont {Rosales-Guzm{\'a}n}\ \emph {et~al.}(2018)\citenamefont {Rosales-Guzm{\'a}n}, \citenamefont {Ndagano},\ and\ \citenamefont {Forbes}}]{Rosales2018}%
  \BibitemOpen
  \bibfield  {author} {\bibinfo {author} {\bibfnamefont {C.}~\bibnamefont {Rosales-Guzm{\'a}n}}, \bibinfo {author} {\bibfnamefont {B.}~\bibnamefont {Ndagano}},\ and\ \bibinfo {author} {\bibfnamefont {A.}~\bibnamefont {Forbes}},\ }\bibfield  {title} {\bibinfo {title} {A review of complex vector light fields and their applications},\ }\href@noop {} {\bibfield  {journal} {\bibinfo  {journal} {Journal of Optics}\ }\textbf {\bibinfo {volume} {20}},\ \bibinfo {pages} {123001} (\bibinfo {year} {2018})}\BibitemShut {NoStop}%
\bibitem [{\citenamefont {Zhan}(2009)}]{Zhan2009}%
  \BibitemOpen
  \bibfield  {author} {\bibinfo {author} {\bibfnamefont {Q.}~\bibnamefont {Zhan}},\ }\bibfield  {title} {\bibinfo {title} {Cylindrical vector beams: from mathematical concepts to applications},\ }\href {https://doi.org/10.1364/AOP.1.000001} {\bibfield  {journal} {\bibinfo  {journal} {Advances in Optics and Photonics}\ }\textbf {\bibinfo {volume} {1}},\ \bibinfo {pages} {1} (\bibinfo {year} {2009})}\BibitemShut {NoStop}%
\bibitem [{\citenamefont {Ndagano}\ \emph {et~al.}(2018)\citenamefont {Ndagano}, \citenamefont {Nape}, \citenamefont {Cox}, \citenamefont {{Rosales-Guzman}},\ and\ \citenamefont {Forbes}}]{Ndagano2018}%
  \BibitemOpen
  \bibfield  {author} {\bibinfo {author} {\bibfnamefont {B.}~\bibnamefont {Ndagano}}, \bibinfo {author} {\bibfnamefont {I.}~\bibnamefont {Nape}}, \bibinfo {author} {\bibfnamefont {M.~A.}\ \bibnamefont {Cox}}, \bibinfo {author} {\bibfnamefont {C.}~\bibnamefont {{Rosales-Guzman}}},\ and\ \bibinfo {author} {\bibfnamefont {A.}~\bibnamefont {Forbes}},\ }\bibfield  {title} {\bibinfo {title} {Creation and {{Detection}} of {{Vector Vortex Modes}} for {{Classical}} and {{Quantum Communication}}},\ }\href {https://doi.org/10.1109/JLT.2017.2766760} {\bibfield  {journal} {\bibinfo  {journal} {Journal of Lightwave Technology}\ }\textbf {\bibinfo {volume} {36}},\ \bibinfo {pages} {292} (\bibinfo {year} {2018})}\BibitemShut {NoStop}%
\bibitem [{\citenamefont {Borges}\ \emph {et~al.}(2010)\citenamefont {Borges}, \citenamefont {Hor-Meyll}, \citenamefont {Huguenin},\ and\ \citenamefont {Khoury}}]{Borges2010}%
  \BibitemOpen
  \bibfield  {author} {\bibinfo {author} {\bibfnamefont {C.}~\bibnamefont {Borges}}, \bibinfo {author} {\bibfnamefont {M.}~\bibnamefont {Hor-Meyll}}, \bibinfo {author} {\bibfnamefont {J.}~\bibnamefont {Huguenin}},\ and\ \bibinfo {author} {\bibfnamefont {A.}~\bibnamefont {Khoury}},\ }\bibfield  {title} {\bibinfo {title} {Bell-like inequality for the spin-orbit separability of a laser beam},\ }\href@noop {} {\bibfield  {journal} {\bibinfo  {journal} {Physical Review A}\ }\textbf {\bibinfo {volume} {82}},\ \bibinfo {pages} {033833} (\bibinfo {year} {2010})}\BibitemShut {NoStop}%
\bibitem [{\citenamefont {Pereira}\ \emph {et~al.}(2014)\citenamefont {Pereira}, \citenamefont {Khoury},\ and\ \citenamefont {Dechoum}}]{Pereira2014}%
  \BibitemOpen
  \bibfield  {author} {\bibinfo {author} {\bibfnamefont {L.}~\bibnamefont {Pereira}}, \bibinfo {author} {\bibfnamefont {A.}~\bibnamefont {Khoury}},\ and\ \bibinfo {author} {\bibfnamefont {K.}~\bibnamefont {Dechoum}},\ }\bibfield  {title} {\bibinfo {title} {Quantum and classical separability of spin-orbit laser modes},\ }\href@noop {} {\bibfield  {journal} {\bibinfo  {journal} {Physical Review A}\ }\textbf {\bibinfo {volume} {90}},\ \bibinfo {pages} {053842} (\bibinfo {year} {2014})}\BibitemShut {NoStop}%
\bibitem [{\citenamefont {D'Ambrosio}\ \emph {et~al.}(2012)\citenamefont {D'Ambrosio}, \citenamefont {Nagali}, \citenamefont {Walborn}, \citenamefont {Aolita}, \citenamefont {Slussarenko}, \citenamefont {Marrucci},\ and\ \citenamefont {Sciarrino}}]{Dambrosio2012}%
  \BibitemOpen
  \bibfield  {author} {\bibinfo {author} {\bibfnamefont {V.}~\bibnamefont {D'Ambrosio}}, \bibinfo {author} {\bibfnamefont {E.}~\bibnamefont {Nagali}}, \bibinfo {author} {\bibfnamefont {S.~P.}\ \bibnamefont {Walborn}}, \bibinfo {author} {\bibfnamefont {L.}~\bibnamefont {Aolita}}, \bibinfo {author} {\bibfnamefont {S.}~\bibnamefont {Slussarenko}}, \bibinfo {author} {\bibfnamefont {L.}~\bibnamefont {Marrucci}},\ and\ \bibinfo {author} {\bibfnamefont {F.}~\bibnamefont {Sciarrino}},\ }\bibfield  {title} {\bibinfo {title} {Complete experimental toolbox for alignment-free quantum communication},\ }\href {https://doi.org/10.1038/ncomms1951} {\bibfield  {journal} {\bibinfo  {journal} {Nature Communications}\ }\textbf {\bibinfo {volume} {3}},\ \bibinfo {pages} {961} (\bibinfo {year} {2012})}\BibitemShut {NoStop}%
\bibitem [{\citenamefont {D'Ambrosio}\ \emph {et~al.}(2016)\citenamefont {D'Ambrosio}, \citenamefont {Carvacho}, \citenamefont {Graffitti}, \citenamefont {Vitelli}, \citenamefont {Piccirillo}, \citenamefont {Marrucci},\ and\ \citenamefont {Sciarrino}}]{Ambrosio2016}%
  \BibitemOpen
  \bibfield  {author} {\bibinfo {author} {\bibfnamefont {V.}~\bibnamefont {D'Ambrosio}}, \bibinfo {author} {\bibfnamefont {G.}~\bibnamefont {Carvacho}}, \bibinfo {author} {\bibfnamefont {F.}~\bibnamefont {Graffitti}}, \bibinfo {author} {\bibfnamefont {C.}~\bibnamefont {Vitelli}}, \bibinfo {author} {\bibfnamefont {B.}~\bibnamefont {Piccirillo}}, \bibinfo {author} {\bibfnamefont {L.}~\bibnamefont {Marrucci}},\ and\ \bibinfo {author} {\bibfnamefont {F.}~\bibnamefont {Sciarrino}},\ }\bibfield  {title} {\bibinfo {title} {Entangled vector vortex beams},\ }\href@noop {} {\bibfield  {journal} {\bibinfo  {journal} {Physical Review A}\ }\textbf {\bibinfo {volume} {94}},\ \bibinfo {pages} {030304} (\bibinfo {year} {2016})}\BibitemShut {NoStop}%
\bibitem [{\citenamefont {Abouraddy}\ and\ \citenamefont {Toussaint}(2006)}]{Abouraddy2006}%
  \BibitemOpen
  \bibfield  {author} {\bibinfo {author} {\bibfnamefont {A.~F.}\ \bibnamefont {Abouraddy}}\ and\ \bibinfo {author} {\bibfnamefont {K.~C.}\ \bibnamefont {Toussaint}},\ }\bibfield  {title} {\bibinfo {title} {Three-{{Dimensional Polarization Control}} in {{Microscopy}}},\ }\href {https://doi.org/10.1103/PhysRevLett.96.153901} {\bibfield  {journal} {\bibinfo  {journal} {Physical Review Letters}\ }\textbf {\bibinfo {volume} {96}},\ \bibinfo {pages} {153901} (\bibinfo {year} {2006})}\BibitemShut {NoStop}%
\bibitem [{\citenamefont {Roxworthy}\ and\ \citenamefont {Toussaint}(2010)}]{Roxworthy2010}%
  \BibitemOpen
  \bibfield  {author} {\bibinfo {author} {\bibfnamefont {B.~J.}\ \bibnamefont {Roxworthy}}\ and\ \bibinfo {author} {\bibfnamefont {K.~C.}\ \bibnamefont {Toussaint}},\ }\bibfield  {title} {\bibinfo {title} {Optical trapping with {$\pi$}-phase cylindrical vector beams},\ }\href {https://doi.org/10.1088/1367-2630/12/7/073012} {\bibfield  {journal} {\bibinfo  {journal} {New Journal of Physics}\ }\textbf {\bibinfo {volume} {12}},\ \bibinfo {pages} {073012} (\bibinfo {year} {2010})}\BibitemShut {NoStop}%
\bibitem [{\citenamefont {Fatemi}(2011)}]{Fatemi2011}%
  \BibitemOpen
  \bibfield  {author} {\bibinfo {author} {\bibfnamefont {F.~K.}\ \bibnamefont {Fatemi}},\ }\bibfield  {title} {\bibinfo {title} {Cylindrical vector beams for rapid polarization-dependent measurements in atomic systems},\ }\href {https://doi.org/10.1364/OE.19.025143} {\bibfield  {journal} {\bibinfo  {journal} {Optics Express}\ }\textbf {\bibinfo {volume} {19}},\ \bibinfo {pages} {25143} (\bibinfo {year} {2011})}\BibitemShut {NoStop}%
\bibitem [{\citenamefont {Neugebauer}\ \emph {et~al.}(2014)\citenamefont {Neugebauer}, \citenamefont {Bauer}, \citenamefont {Banzer},\ and\ \citenamefont {Leuchs}}]{Neugebauer2014}%
  \BibitemOpen
  \bibfield  {author} {\bibinfo {author} {\bibfnamefont {M.}~\bibnamefont {Neugebauer}}, \bibinfo {author} {\bibfnamefont {T.}~\bibnamefont {Bauer}}, \bibinfo {author} {\bibfnamefont {P.}~\bibnamefont {Banzer}},\ and\ \bibinfo {author} {\bibfnamefont {G.}~\bibnamefont {Leuchs}},\ }\bibfield  {title} {\bibinfo {title} {Polarization {{Tailored Light Driven Directional Optical Nanobeacon}}},\ }\href {https://doi.org/10.1021/nl5003526} {\bibfield  {journal} {\bibinfo  {journal} {Nano Letters}\ }\textbf {\bibinfo {volume} {14}},\ \bibinfo {pages} {2546} (\bibinfo {year} {2014})}\BibitemShut {NoStop}%
\bibitem [{\citenamefont {Parigi}\ \emph {et~al.}(2015)\citenamefont {Parigi}, \citenamefont {D'Ambrosio}, \citenamefont {Arnold}, \citenamefont {Marrucci}, \citenamefont {Sciarrino},\ and\ \citenamefont {Laurat}}]{Parigi2015}%
  \BibitemOpen
  \bibfield  {author} {\bibinfo {author} {\bibfnamefont {V.}~\bibnamefont {Parigi}}, \bibinfo {author} {\bibfnamefont {V.}~\bibnamefont {D'Ambrosio}}, \bibinfo {author} {\bibfnamefont {C.}~\bibnamefont {Arnold}}, \bibinfo {author} {\bibfnamefont {L.}~\bibnamefont {Marrucci}}, \bibinfo {author} {\bibfnamefont {F.}~\bibnamefont {Sciarrino}},\ and\ \bibinfo {author} {\bibfnamefont {J.}~\bibnamefont {Laurat}},\ }\bibfield  {title} {\bibinfo {title} {Storage and retrieval of vector beams of light in a multiple-degree-of-freedom quantum memory},\ }\href {https://doi.org/10.1038/ncomms8706} {\bibfield  {journal} {\bibinfo  {journal} {Nature Communications}\ }\textbf {\bibinfo {volume} {6}},\ \bibinfo {pages} {7706} (\bibinfo {year} {2015})}\BibitemShut {NoStop}%
\bibitem [{\citenamefont {Milione}\ \emph {et~al.}(2015{\natexlab{a}})\citenamefont {Milione}, \citenamefont {Nguyen}, \citenamefont {Leach}, \citenamefont {Nolan},\ and\ \citenamefont {Alfano}}]{Milione2015b}%
  \BibitemOpen
  \bibfield  {author} {\bibinfo {author} {\bibfnamefont {G.}~\bibnamefont {Milione}}, \bibinfo {author} {\bibfnamefont {T.~A.}\ \bibnamefont {Nguyen}}, \bibinfo {author} {\bibfnamefont {J.}~\bibnamefont {Leach}}, \bibinfo {author} {\bibfnamefont {D.~A.}\ \bibnamefont {Nolan}},\ and\ \bibinfo {author} {\bibfnamefont {R.~R.}\ \bibnamefont {Alfano}},\ }\bibfield  {title} {\bibinfo {title} {Using the nonseparability of vector beams to encode information for optical communication},\ }\href@noop {} {\bibfield  {journal} {\bibinfo  {journal} {Optics letters}\ }\textbf {\bibinfo {volume} {40}},\ \bibinfo {pages} {4887} (\bibinfo {year} {2015}{\natexlab{a}})}\BibitemShut {NoStop}%
\bibitem [{\citenamefont {Yuan}\ \emph {et~al.}(2022)\citenamefont {Yuan}, \citenamefont {Xiao}, \citenamefont {Liu}, \citenamefont {Fu}, \citenamefont {Qu}, \citenamefont {Gbur},\ and\ \citenamefont {Cai}}]{yuan2022}%
  \BibitemOpen
  \bibfield  {author} {\bibinfo {author} {\bibfnamefont {Y.}~\bibnamefont {Yuan}}, \bibinfo {author} {\bibfnamefont {X.}~\bibnamefont {Xiao}}, \bibinfo {author} {\bibfnamefont {D.}~\bibnamefont {Liu}}, \bibinfo {author} {\bibfnamefont {P.}~\bibnamefont {Fu}}, \bibinfo {author} {\bibfnamefont {J.}~\bibnamefont {Qu}}, \bibinfo {author} {\bibfnamefont {G.}~\bibnamefont {Gbur}},\ and\ \bibinfo {author} {\bibfnamefont {Y.}~\bibnamefont {Cai}},\ }\bibfield  {title} {\bibinfo {title} {Mitigating orbital angular momentum crosstalk in an optical communication uplink channel using cylindrical vector beams},\ }\href {https://doi.org/10.1080/17455030.2022.2053609} {\bibfield  {journal} {\bibinfo  {journal} {Waves in Random and Complex Media}\ }\textbf {\bibinfo {volume} {0}},\ \bibinfo {pages} {1} (\bibinfo {year} {2022})}\BibitemShut {NoStop}%
\bibitem [{\citenamefont {Cheng}\ \emph {et~al.}(2023)\citenamefont {Cheng}, \citenamefont {Dong}, \citenamefont {Shi}, \citenamefont {Mohammed}, \citenamefont {Guo}, \citenamefont {Yi}, \citenamefont {Wang},\ and\ \citenamefont {Li}}]{Cheng2023}%
  \BibitemOpen
  \bibfield  {author} {\bibinfo {author} {\bibfnamefont {M.}~\bibnamefont {Cheng}}, \bibinfo {author} {\bibfnamefont {K.}~\bibnamefont {Dong}}, \bibinfo {author} {\bibfnamefont {C.}~\bibnamefont {Shi}}, \bibinfo {author} {\bibfnamefont {A.-A. H.~T.}\ \bibnamefont {Mohammed}}, \bibinfo {author} {\bibfnamefont {L.}~\bibnamefont {Guo}}, \bibinfo {author} {\bibfnamefont {X.}~\bibnamefont {Yi}}, \bibinfo {author} {\bibfnamefont {P.}~\bibnamefont {Wang}},\ and\ \bibinfo {author} {\bibfnamefont {J.}~\bibnamefont {Li}},\ }\bibfield  {title} {\bibinfo {title} {Enhancing {{Performance}} of {{Air}}\textendash{{Ground OAM Communication System Utilizing Vector Vortex Beams}} in the {{Atmosphere}}},\ }\href {https://doi.org/10.3390/photonics10010041} {\bibfield  {journal} {\bibinfo  {journal} {Photonics}\ }\textbf {\bibinfo {volume} {10}},\ \bibinfo {pages} {41} (\bibinfo {year} {2023})}\BibitemShut {NoStop}%
\bibitem [{\citenamefont {Cheng}\ \emph {et~al.}(2009)\citenamefont {Cheng}, \citenamefont {Haus},\ and\ \citenamefont {Zhan}}]{cheng2009}%
  \BibitemOpen
  \bibfield  {author} {\bibinfo {author} {\bibfnamefont {W.}~\bibnamefont {Cheng}}, \bibinfo {author} {\bibfnamefont {J.~W.}\ \bibnamefont {Haus}},\ and\ \bibinfo {author} {\bibfnamefont {Q.}~\bibnamefont {Zhan}},\ }\bibfield  {title} {\bibinfo {title} {Propagation of vector vortex beams through a turbulent atmosphere},\ }\href {https://doi.org/10.1364/OE.17.017829} {\bibfield  {journal} {\bibinfo  {journal} {Optics Express}\ }\textbf {\bibinfo {volume} {17}},\ \bibinfo {pages} {17829} (\bibinfo {year} {2009})}\BibitemShut {NoStop}%
\bibitem [{\citenamefont {Rubano}\ \emph {et~al.}(2019)\citenamefont {Rubano}, \citenamefont {Cardano}, \citenamefont {Piccirillo},\ and\ \citenamefont {Marrucci}}]{Rubano2019}%
  \BibitemOpen
  \bibfield  {author} {\bibinfo {author} {\bibfnamefont {A.}~\bibnamefont {Rubano}}, \bibinfo {author} {\bibfnamefont {F.}~\bibnamefont {Cardano}}, \bibinfo {author} {\bibfnamefont {B.}~\bibnamefont {Piccirillo}},\ and\ \bibinfo {author} {\bibfnamefont {L.}~\bibnamefont {Marrucci}},\ }\bibfield  {title} {\bibinfo {title} {Q-plate technology: a progress review},\ }\href@noop {} {\bibfield  {journal} {\bibinfo  {journal} {Journal of the optical society of america B}\ }\textbf {\bibinfo {volume} {36}},\ \bibinfo {pages} {D70} (\bibinfo {year} {2019})}\BibitemShut {NoStop}%
\bibitem [{\citenamefont {Milione}\ \emph {et~al.}(2015{\natexlab{b}})\citenamefont {Milione}, \citenamefont {Lavery}, \citenamefont {Huang}, \citenamefont {Ren}, \citenamefont {Xie}, \citenamefont {Nguyen}, \citenamefont {Karimi}, \citenamefont {Marrucci}, \citenamefont {Nolan}, \citenamefont {Alfano} \emph {et~al.}}]{Milione2015}%
  \BibitemOpen
  \bibfield  {author} {\bibinfo {author} {\bibfnamefont {G.}~\bibnamefont {Milione}}, \bibinfo {author} {\bibfnamefont {M.~P.}\ \bibnamefont {Lavery}}, \bibinfo {author} {\bibfnamefont {H.}~\bibnamefont {Huang}}, \bibinfo {author} {\bibfnamefont {Y.}~\bibnamefont {Ren}}, \bibinfo {author} {\bibfnamefont {G.}~\bibnamefont {Xie}}, \bibinfo {author} {\bibfnamefont {T.~A.}\ \bibnamefont {Nguyen}}, \bibinfo {author} {\bibfnamefont {E.}~\bibnamefont {Karimi}}, \bibinfo {author} {\bibfnamefont {L.}~\bibnamefont {Marrucci}}, \bibinfo {author} {\bibfnamefont {D.~A.}\ \bibnamefont {Nolan}}, \bibinfo {author} {\bibfnamefont {R.~R.}\ \bibnamefont {Alfano}}, \emph {et~al.},\ }\bibfield  {title} {\bibinfo {title} {4$\times$ 20 gbit/s mode division multiplexing over free space using vector modes and aq-plate mode (de) multiplexer},\ }\href@noop {} {\bibfield  {journal} {\bibinfo  {journal} {Optics letters}\ }\textbf {\bibinfo {volume} {40}},\ \bibinfo {pages} {1980} (\bibinfo {year} {2015}{\natexlab{b}})}\BibitemShut
  {NoStop}%
\bibitem [{\citenamefont {Jia}\ \emph {et~al.}(2019)\citenamefont {Jia}, \citenamefont {Chang}, \citenamefont {Yang}, \citenamefont {Liu}, \citenamefont {Wang}, \citenamefont {Gao}, \citenamefont {Li},\ and\ \citenamefont {Zhang}}]{Jia2019}%
  \BibitemOpen
  \bibfield  {author} {\bibinfo {author} {\bibfnamefont {J.}~\bibnamefont {Jia}}, \bibinfo {author} {\bibfnamefont {Z.}~\bibnamefont {Chang}}, \bibinfo {author} {\bibfnamefont {H.}~\bibnamefont {Yang}}, \bibinfo {author} {\bibfnamefont {Q.}~\bibnamefont {Liu}}, \bibinfo {author} {\bibfnamefont {F.}~\bibnamefont {Wang}}, \bibinfo {author} {\bibfnamefont {H.}~\bibnamefont {Gao}}, \bibinfo {author} {\bibfnamefont {F.}~\bibnamefont {Li}},\ and\ \bibinfo {author} {\bibfnamefont {P.}~\bibnamefont {Zhang}},\ }\bibfield  {title} {\bibinfo {title} {Mode sorter designed for (de) multiplexing vector vortex modes},\ }\href@noop {} {\bibfield  {journal} {\bibinfo  {journal} {Applied Optics}\ }\textbf {\bibinfo {volume} {58}},\ \bibinfo {pages} {7094} (\bibinfo {year} {2019})}\BibitemShut {NoStop}%
\bibitem [{\citenamefont {dos Santos}\ \emph {et~al.}(2021)\citenamefont {dos Santos}, \citenamefont {Salles}, \citenamefont {Damaceno}, \citenamefont {Menezes}, \citenamefont {Corso}, \citenamefont {Martinelli}, \citenamefont {Ribeiro},\ and\ \citenamefont {de~Ara{\'u}jo}}]{Santos2021}%
  \BibitemOpen
  \bibfield  {author} {\bibinfo {author} {\bibfnamefont {G.~H.}\ \bibnamefont {dos Santos}}, \bibinfo {author} {\bibfnamefont {D.~C.}\ \bibnamefont {Salles}}, \bibinfo {author} {\bibfnamefont {M.~G.}\ \bibnamefont {Damaceno}}, \bibinfo {author} {\bibfnamefont {B.~T.~d.}\ \bibnamefont {Menezes}}, \bibinfo {author} {\bibfnamefont {C.}~\bibnamefont {Corso}}, \bibinfo {author} {\bibfnamefont {M.}~\bibnamefont {Martinelli}}, \bibinfo {author} {\bibfnamefont {P.~S.}\ \bibnamefont {Ribeiro}},\ and\ \bibinfo {author} {\bibfnamefont {R.~M.}\ \bibnamefont {de~Ara{\'u}jo}},\ }\bibfield  {title} {\bibinfo {title} {Decomposing spatial mode superpositions with a triangular optical cavity},\ }\href@noop {} {\bibfield  {journal} {\bibinfo  {journal} {Physical Review Applied}\ }\textbf {\bibinfo {volume} {16}},\ \bibinfo {pages} {034008} (\bibinfo {year} {2021})}\BibitemShut {NoStop}%
\bibitem [{\citenamefont {Rodrigues}\ \emph {et~al.}(2024)\citenamefont {Rodrigues}, \citenamefont {Marques~Fagundes}, \citenamefont {Salles}, \citenamefont {dos Santos}, \citenamefont {Kondo}, \citenamefont {Souto~Ribeiro}, \citenamefont {Khoury},\ and\ \citenamefont {Medeiros~de Ara{\'u}jo}}]{Rodrigues2024}%
  \BibitemOpen
  \bibfield  {author} {\bibinfo {author} {\bibfnamefont {L.~M.}\ \bibnamefont {Rodrigues}}, \bibinfo {author} {\bibfnamefont {L.}~\bibnamefont {Marques~Fagundes}}, \bibinfo {author} {\bibfnamefont {D.}~\bibnamefont {Salles}}, \bibinfo {author} {\bibfnamefont {G.~H.}\ \bibnamefont {dos Santos}}, \bibinfo {author} {\bibfnamefont {J.}~\bibnamefont {Kondo}}, \bibinfo {author} {\bibfnamefont {P.}~\bibnamefont {Souto~Ribeiro}}, \bibinfo {author} {\bibfnamefont {A.}~\bibnamefont {Khoury}},\ and\ \bibinfo {author} {\bibfnamefont {R.}~\bibnamefont {Medeiros~de Ara{\'u}jo}},\ }\bibfield  {title} {\bibinfo {title} {Resonance of vector vortex beams in a triangular optical cavity},\ }\href@noop {} {\bibfield  {journal} {\bibinfo  {journal} {Scientific Reports}\ }\textbf {\bibinfo {volume} {14}},\ \bibinfo {pages} {10542} (\bibinfo {year} {2024})}\BibitemShut {NoStop}%
\bibitem [{\citenamefont {Marques~Fagundes}\ \emph {et~al.}(2025)\citenamefont {Marques~Fagundes}, \citenamefont {Souto~Ribeiro},\ and\ \citenamefont {Medeiros~de Araújo}}]{Fagundes2025}%
  \BibitemOpen
  \bibfield  {author} {\bibinfo {author} {\bibfnamefont {L.}~\bibnamefont {Marques~Fagundes}}, \bibinfo {author} {\bibfnamefont {P.~H.}\ \bibnamefont {Souto~Ribeiro}},\ and\ \bibinfo {author} {\bibfnamefont {R.}~\bibnamefont {Medeiros~de Araújo}},\ }\bibfield  {title} {\bibinfo {title} {Discrimination of vortex and pseudovortex beams with a triangular optical cavity},\ }\href@noop {} {\bibfield  {journal} {\bibinfo  {journal} {Journal of Optics}\ } (\bibinfo {year} {2025})}\BibitemShut {NoStop}%
\bibitem [{\citenamefont {Maurer}\ \emph {et~al.}(2007)\citenamefont {Maurer}, \citenamefont {Jesacher}, \citenamefont {F{\"u}rhapter}, \citenamefont {Bernet},\ and\ \citenamefont {Ritsch-Marte}}]{Maurer2007}%
  \BibitemOpen
  \bibfield  {author} {\bibinfo {author} {\bibfnamefont {C.}~\bibnamefont {Maurer}}, \bibinfo {author} {\bibfnamefont {A.}~\bibnamefont {Jesacher}}, \bibinfo {author} {\bibfnamefont {S.}~\bibnamefont {F{\"u}rhapter}}, \bibinfo {author} {\bibfnamefont {S.}~\bibnamefont {Bernet}},\ and\ \bibinfo {author} {\bibfnamefont {M.}~\bibnamefont {Ritsch-Marte}},\ }\bibfield  {title} {\bibinfo {title} {Tailoring of arbitrary optical vector beams},\ }\href@noop {} {\bibfield  {journal} {\bibinfo  {journal} {New Journal of Physics}\ }\textbf {\bibinfo {volume} {9}},\ \bibinfo {pages} {78} (\bibinfo {year} {2007})}\BibitemShut {NoStop}%
\bibitem [{\citenamefont {McWilliam}\ \emph {et~al.}(2022)\citenamefont {McWilliam}, \citenamefont {Cisowski}, \citenamefont {Bennett},\ and\ \citenamefont {Franke-Arnold}}]{McWilliam2022}%
  \BibitemOpen
  \bibfield  {author} {\bibinfo {author} {\bibfnamefont {A.}~\bibnamefont {McWilliam}}, \bibinfo {author} {\bibfnamefont {C.~M.}\ \bibnamefont {Cisowski}}, \bibinfo {author} {\bibfnamefont {R.}~\bibnamefont {Bennett}},\ and\ \bibinfo {author} {\bibfnamefont {S.}~\bibnamefont {Franke-Arnold}},\ }\bibfield  {title} {\bibinfo {title} {Angular momentum redirection phase of vector beams in a non-planar geometry},\ }\href@noop {} {\bibfield  {journal} {\bibinfo  {journal} {Nanophotonics}\ }\textbf {\bibinfo {volume} {11}},\ \bibinfo {pages} {727} (\bibinfo {year} {2022})}\BibitemShut {NoStop}%
\bibitem [{\citenamefont {Karan}\ \emph {et~al.}(2022)\citenamefont {Karan}, \citenamefont {Ruchi}, \citenamefont {Mohta},\ and\ \citenamefont {Jha}}]{Karan2022}%
  \BibitemOpen
  \bibfield  {author} {\bibinfo {author} {\bibfnamefont {S.}~\bibnamefont {Karan}}, \bibinfo {author} {\bibnamefont {Ruchi}}, \bibinfo {author} {\bibfnamefont {P.}~\bibnamefont {Mohta}},\ and\ \bibinfo {author} {\bibfnamefont {A.~K.}\ \bibnamefont {Jha}},\ }\bibfield  {title} {\bibinfo {title} {Quantifying polarization changes induced by rotating dove prisms and k-mirrors},\ }\href@noop {} {\bibfield  {journal} {\bibinfo  {journal} {Applied Optics}\ }\textbf {\bibinfo {volume} {61}},\ \bibinfo {pages} {8302} (\bibinfo {year} {2022})}\BibitemShut {NoStop}%
\bibitem [{\citenamefont {Karan}\ \emph {et~al.}(2024)\citenamefont {Karan}, \citenamefont {Senapati},\ and\ \citenamefont {Jha}}]{Karan2024}%
  \BibitemOpen
  \bibfield  {author} {\bibinfo {author} {\bibfnamefont {S.}~\bibnamefont {Karan}}, \bibinfo {author} {\bibfnamefont {N.}~\bibnamefont {Senapati}},\ and\ \bibinfo {author} {\bibfnamefont {A.~K.}\ \bibnamefont {Jha}},\ }\bibfield  {title} {\bibinfo {title} {Wavefront rotator with near-zero mean polarization change},\ }\href@noop {} {\bibfield  {journal} {\bibinfo  {journal} {Applied Optics}\ }\textbf {\bibinfo {volume} {63}},\ \bibinfo {pages} {4552} (\bibinfo {year} {2024})}\BibitemShut {NoStop}%
\bibitem [{\citenamefont {Lacerda}(2025)}]{code_Kmirror_2025}%
  \BibitemOpen
  \bibfield  {author} {\bibinfo {author} {\bibfnamefont {P.~S.}\ \bibnamefont {Lacerda}},\ }\href@noop {} {\bibinfo {title} {K-mirror mean polarization change}},\ \bibinfo {howpublished} {\url{https://github.com/Grock42/K-mirror}} (\bibinfo {year} {2025}),\ \bibinfo {note} {commit 9e7f86f.}\BibitemShut {Stop}%
\bibitem [{\citenamefont {Milione}\ \emph {et~al.}(2011)\citenamefont {Milione}, \citenamefont {Sztul}, \citenamefont {Nolan},\ and\ \citenamefont {Alfano}}]{Milione2011}%
  \BibitemOpen
  \bibfield  {author} {\bibinfo {author} {\bibfnamefont {G.}~\bibnamefont {Milione}}, \bibinfo {author} {\bibfnamefont {H.}~\bibnamefont {Sztul}}, \bibinfo {author} {\bibfnamefont {D.}~\bibnamefont {Nolan}},\ and\ \bibinfo {author} {\bibfnamefont {R.}~\bibnamefont {Alfano}},\ }\bibfield  {title} {\bibinfo {title} {Higher-order poincar{\'e} sphere, stokes parameters, and the angular momentum of light},\ }\href@noop {} {\bibfield  {journal} {\bibinfo  {journal} {Physical review letters}\ }\textbf {\bibinfo {volume} {107}},\ \bibinfo {pages} {053601} (\bibinfo {year} {2011})}\BibitemShut {NoStop}%
\bibitem [{\citenamefont {Gouy}(1890)}]{Gouy1890}%
  \BibitemOpen
  \bibfield  {author} {\bibinfo {author} {\bibfnamefont {L.~G.}\ \bibnamefont {Gouy}},\ }\href@noop {} {\emph {\bibinfo {title} {Sur une propri{\'e}t{\'e} nouvelle des ondes lumineuses}}}\ (\bibinfo  {publisher} {Gauthier-Villars},\ \bibinfo {address} {Paris},\ \bibinfo {year} {1890})\BibitemShut {NoStop}%
\bibitem [{\citenamefont {Kogelnik}\ and\ \citenamefont {Li}(1966)}]{Kogelnik1966}%
  \BibitemOpen
  \bibfield  {author} {\bibinfo {author} {\bibfnamefont {H.}~\bibnamefont {Kogelnik}}\ and\ \bibinfo {author} {\bibfnamefont {T.}~\bibnamefont {Li}},\ }\bibfield  {title} {\bibinfo {title} {Laser beams and resonators},\ }\href@noop {} {\bibfield  {journal} {\bibinfo  {journal} {Applied optics}\ }\textbf {\bibinfo {volume} {5}},\ \bibinfo {pages} {1550} (\bibinfo {year} {1966})}\BibitemShut {NoStop}%
\end{thebibliography}
\end{document}